\documentclass[prl,twocolumn,superscriptaddress,showpacs]{revtex4}
\usepackage{graphicx,amsmath}

\newcommand{\Eq}[1]{Eq.~\eqref{#1}}
\newcommand{\eq}[1]{\eqref{#1}}

\newcommand{\beq}{\begin{equation}}
\newcommand{\eeq}{\end{equation}}

\newcommand{\beqa}{\begin{eqnarray}}
\newcommand{\eeqa}{\end{eqnarray}}

\newcommand{\Beqa}{\begin{eqnarray*}}
\newcommand{\Eeqa}{\end{eqnarray*}}
\newcommand{\nn}{\nonumber}
\newcommand{\ds}{\displaystyle}

\newcommand{\pdag}{{\phantom{\dagger}}}

\DeclareMathOperator{\sign}{sgn}

\DeclareMathOperator{\tsum}{\textstyle\sum}

\newcommand{\PRL}[3]{Phys. Rev. Lett.~\textbf{#1}, #2 (#3)}
\newcommand{\PRB}[3]{Phys. Rev. B~\textbf{#1}, #2 (#3)}
\newcommand{\PR}[3]{Phys. Rev.~\textbf{#1}, #2 (#3)}

\newcommand{\JETP}[3]{Sov. Phys. JETP~\textbf{#1}, #2 (#3)}
\newcommand{\ZhETF}[3]{Zh. Eksp. Teor. Fiz.~\textbf{#1}, #2 (#3)}
\newcommand{\JPCM}[3]{J. Phys. Cond. Matt.~\textbf{#1}, #2 (#3)}

\newcommand{\JMP}[3]{J. Math. Phys.~\textbf{#1}, #2 (#3)}

\newcommand{\cm}[1]{cond-mat/{#1}}

\newcommand{\etal}{\textit{et al.}}

\begin{document}

\title{Dynamic response of one-dimensional interacting fermions}

\author{M. Pustilnik}
\affiliation{School of Physics, Georgia Institute of Technology, Atlanta, GA 30332}
\author{M. Khodas}
\affiliation{William I. Fine Theoretical Physics Institute and School
 of Physics and Astronomy, University of Minnesota, Minneapolis, MN 55455}
\author{A. Kamenev}
\affiliation{William I. Fine Theoretical Physics Institute and School
 of Physics and Astronomy, University of Minnesota, Minneapolis, MN 55455}
\author{L.I. Glazman}
\affiliation{William I. Fine Theoretical Physics Institute and School
 of Physics and Astronomy, University of Minnesota, Minneapolis, MN 55455}

\begin{abstract}
We evaluate the dynamic structure factor $S(q,\omega)$ of interacting 
one-dimensional spinless fermions with a nonlinear dispersion relation. 
The combined effect of the nonlinear dispersion and of the interactions 
leads to new universal features of $S(q,\omega)$. The sharp peak 
$S(q,\omega)\propto q\delta(\omega-uq)$, characteristic for the 
Tomonaga-Luttinger model, broadens up; $S(q,\omega)$ for a fixed $q$ 
becomes finite at arbitrarily large $\omega$.  The main spectral weight, 
however, is confined to a narrow frequency interval of the width
$\delta\omega\sim q^2\!/m$. At the boundaries of this interval the
structure factor exhibits power-law singularities with exponents
depending on the interaction strength and on the wave number $q$.
\end{abstract}

\maketitle

Low-energy properties of fermionic systems are sensitive to interactions 
between fermions. The effect of interactions is the strongest in one 
dimension (1D), where single-particle correlation functions exhibit 
power-law singularities, in a striking departure from the behavior in higher 
dimensions. Much of our current understanding of 1D fermions is based 
on the Tomonaga-Luttinger (TL) model \cite{TL}. The crucial ingredient 
of the model is the assumption of a \textit{strictly linear} fermionic 
dispersion relation. The TL model, often used in conjunction with a 
powerful bosonization technique \cite{Giamarchi}, allows one to evaluate
various correlation functions, such as momentum-resolved \cite{DL,LP}
and local \cite{KF} single-particle densities of states.

Unlike the single-particle correlation functions, the two-particle
correlation functions of the TL model exhibit behavior rather
compatible with that expected for a Fermi liquid with the linear
spectrum of quasiparticles. For example, the dynamic structure factor
(the density-density correlation function) 
\beq S(q,\omega) =
\!\int\!\!dx\,dt\,e^{i(\omega t - qx)}
\bigl\langle\rho(x,t)\rho(0,0)\bigr\rangle
\label{1}
\eeq
at small $q$ takes the form \cite{DL}
$S_\text{TL}(q,\omega)\!\propto q\delta(\omega - uq)$. It means
that the quanta of density waves propagating with plasma velocity
$u$ are true eigenstates of the TL model;
these bosonic excitations have an infinite lifetime.

Below we show that such a simple behavior is an artefact of the linear
spectrum approximation. In reality, the spectrum of fermions always 
has some nonlinearity,
\beq
\xi_{R/L,\,k}\! = \pm\,vk + k^2\!/2m+\ldots,
\label{2}
\eeq
where the upper/lower sign corresponds to the right/left movers
(R/L), and $k=p\mp p_F$ are momenta measured from the Fermi  
points $\pm \,p_F$. (For Galilean-invariant systems the expansion 
\eq{2} terminates at $k^2$.)

The finite curvature ($1/m\neq 0$) affects drastically the functional
form of $S(q,\omega)$. In a clear deviation from the results of TL 
model, power-law singularities now arise not only in the single-particle 
correlation functions, but in the structure factor as well. We show that 
the singularities in these two very different objects have a common 
origin, proliferation of low-energy particle-hole pairs, and evaluate 
the corresponding exponents.

Because of the success of the bosonization technique~\cite{Giamarchi},
it is tempting to treat the spectrum nonlinearity as a weak interaction
between the TL bosons. Indeed, the nonlinearity gives rise to a
three-boson interaction with the coupling constant $\propto 1/m$
\cite{Haldane}. However, attempts to treat this interaction perturbatively 
fail as the finite-order contributions to the boson's self-energy diverge at 
the mass shell \cite{Samokhin}. A reliable method of resumming the 
corresponding series is yet to be developed.

\begin{figure}[h]
\includegraphics[width=0.85\columnwidth]{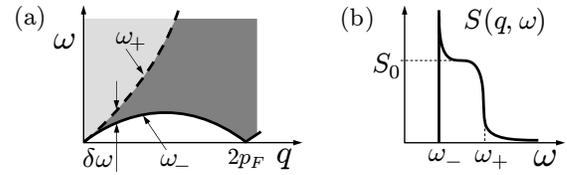}
\caption{
(a) Without interactions, the structure factor differs from zero
only for $\omega_{-}\!\!<\!\omega\!<\!\omega_{+}$. In the
presence of interactions, $S\neq 0$ at $\omega>\omega_+$
as well. For repulsive interactions, $S(q,\omega)$
has a power-law divergence at $\omega=\omega_-(q)$ with 
an exponent depending on $q$.
(b) Sketch of $S(q,\omega)$ at a fixed small $q\ll p_F$, see 
Eqs. \eq{5} and \eq{7}.
\label{Fig1}
}
\end{figure}

We found it more productive to approach the problem from the fermionic
perspective. One has then a benchmark reference point: the structure
factor of free fermions. At zero temperature, the structure
factor has a simple physical meaning of the absorption rate
of a photon with energy $\omega$ and momentum $q$~\cite{fdt}. 
Without interaction, absorption of a photon results in a creation 
of a single particle-hole pair. For a fixed $q<2p_F$, the energy of 
this pair lies within the interval
\beq 
\omega_-<\omega<\omega_+, \quad
\omega_\pm = uq \pm\,q^2\!/2m
\label{3}
\eeq
(for free fermions $u = v$). The bounds $\omega_+$ $(\omega_-)$ 
correspond to particle-hole pairs in which a hole (a particle) is 
created just below (just above) the Fermi energy. If $\omega$ is 
outside the interval \eq{3}, the energy and momentum conservation 
laws can not be satisfied and the structure factor vanishes. Within 
this interval, $S(q,\omega)$ is independent of $\omega$, 
\beq 
S(q,\omega) = S_0(q) = m/q, 
\quad 
\omega_-<\omega<\omega_+, 
\label{4} 
\eeq 
see Fig.~\ref{Fig1}. Accordingly, $S(q,\omega)$ at a fixed $q$
exhibits a peak which has a ``rectangular'' shape with the width
$\delta\omega = \omega_+ - \omega_- = q^2\!/m$~\cite{width}.

If even a weak interaction is now turned on, its effect on the
structure factor is dramatic, see Fig.~\ref{Fig1}. The structure
factor still vanishes below the (renormalized) lower absorption
``edge'' $\omega=\omega_-$, but above the edge $S(q,\omega)$
develops a power-law singularity,
\beq 
\frac{S(q,\omega)}{S_0(q)}
= \left[\frac{\delta\omega~}{\omega-\omega_-\!}\right]^\mu, 
\quad
0<\omega-\omega_-\ll\delta\omega. 
\label{5} 
\eeq 
The exponent $\mu$ here is a smooth function
of $q$, 
\beq 
\mu(q) = \frac{m}{\pi q}\, \bigl(V_0 - V_q\bigr)
\label{6}
\eeq
($V_p$ is the Fourier component of the intra-branch interaction 
potential, see \Eq{10} below). The divergence in \Eq{5} has the 
same origin as the familiar X-ray edge singularity  in metals~\cite{Mahan}.

In the presence of interactions, absorption of a photon is accompanied by
the creation of multiple particle-hole pairs, which allows the conservation
laws to be satisfied at arbitrarily high $\omega$~\cite{Nozieres}, so that
$S(q,\omega)\neq 0$ for $\omega >\omega_+$. However, at
$\omega = \omega_+$ the structure factor still exhibits a power-law
singularity. At $|\omega-\omega_+|\ll\delta\omega$ we find
\beq
\!\frac{S(q,\omega)}{S_0(q)} =\!
\left\lbrace
\begin{array}{lc}
\ds\left[\frac{\omega_+\!-\omega}{\delta\omega}\right]^\mu
\!+ \,\frac{\nu}{\mu}\,,\!
&
\quad\omega<\omega_+
\\ \\
\ds\frac{\nu}{\mu}
\left(
1 - \left[\frac{\omega-\omega_+\!}{\delta\omega~}\right]^{\mu}
\,\right),\!
&
\quad\omega>\omega_+
\end{array}
\right.
\label{7}
\eeq
with
\beq
\nu(q) = \left(\frac{q}{4mu}\right)^2\!
\left(\frac{U_0}{2\pi u}\right)^2\!
\ll\mu(q).
\label{8}
\eeq
Here $U_0$ is the interaction between the right- and left-movers, 
see \Eq{10} below.

Finally, at high frequencies $S(q,\omega)$ can be evaluated~\cite{drag} 
in the second order of perturbation theory in the interaction between fermions,
\beq
 S(q,\omega)=2\nu\, \frac{uq^2}{\omega^2 - u^2 q^2}\,,
\quad
\omega-\omega_+\gg \delta\omega.
\label{9}
\eeq
Note that $S\propto q^4$ in \Eq{9}, as expected for a multipair
contribution to the photon absorption rate \cite{Nozieres}. Therefore,
for $q\ll p_F$ the high-frequency ``tail'' yields a negligible contribution
to the $f$-sum rule: The main spectral weight of $S(q,\omega)$ is still
confined within the narrow frequency window \eq{3}. At the borders
of this interval $S(q,\omega)$ develops power-law singularities, see
Eqs. \eq{5} and \eq{7}. The resulting rather peculiar shape of the
peak in $S(q,\omega)$ is sketched in Fig.~\ref{Fig1}(b). It is certainly
very different from a simple Lorentzian assumed in, e.g., Ref.~\cite{Kopietz}.
If one were to interpret the finite width of the peak as a lifetime
of the TL bosons, one would conclude that the boson's decay is
manifestly non-exponential. Instead, it is governed by power
laws, indicating strong nonlinearity-induced correlations between
the TL bosons.

Equations \eq{5}-\eq{8} represent the main result of this
Letter. We now outline their derivation. 

We describe spinless 1D
fermions by the Hamiltonian \beqa H &=&
\sum_{\alpha,k}\xi_{\alpha,k}^\pdag
\psi_{\alpha,k}^\dagger\psi_{\alpha,k}^\pdag
\label{10}\\
&\,&\,+\,\,\frac{1}{2L}\!\sum_{p\,\neq 0}
\Bigl\lbrace
V_{p} \tsum_\alpha\!\rho_{\alpha,p\,}\rho_{\alpha,-p}
\,+\, 2U_{p\,} \rho_{R,p\,}\rho_{L,-p}
\Bigr\rbrace,
\nn
\eeqa
where $\alpha = R,L$, the single-particle energies $\xi_{\alpha,k}$ are
given by \Eq{2},
$\rho_{\alpha,p} = \sum_k \psi_{\alpha,k-p}^\dagger\psi_{\alpha,k}^\pdag$,
and $L$ is size of the system. We consider short-range interaction, 
so that $V_0,U_0$ are finite and $V_0-V_p\propto p^2$ (and 
similarly for $U_p$) for small $p\ll p_F$. (Note that $V_p=U_p$ 
for Galilean-invariant systems).

\begin{figure}[h]
\includegraphics[width=0.65\columnwidth]{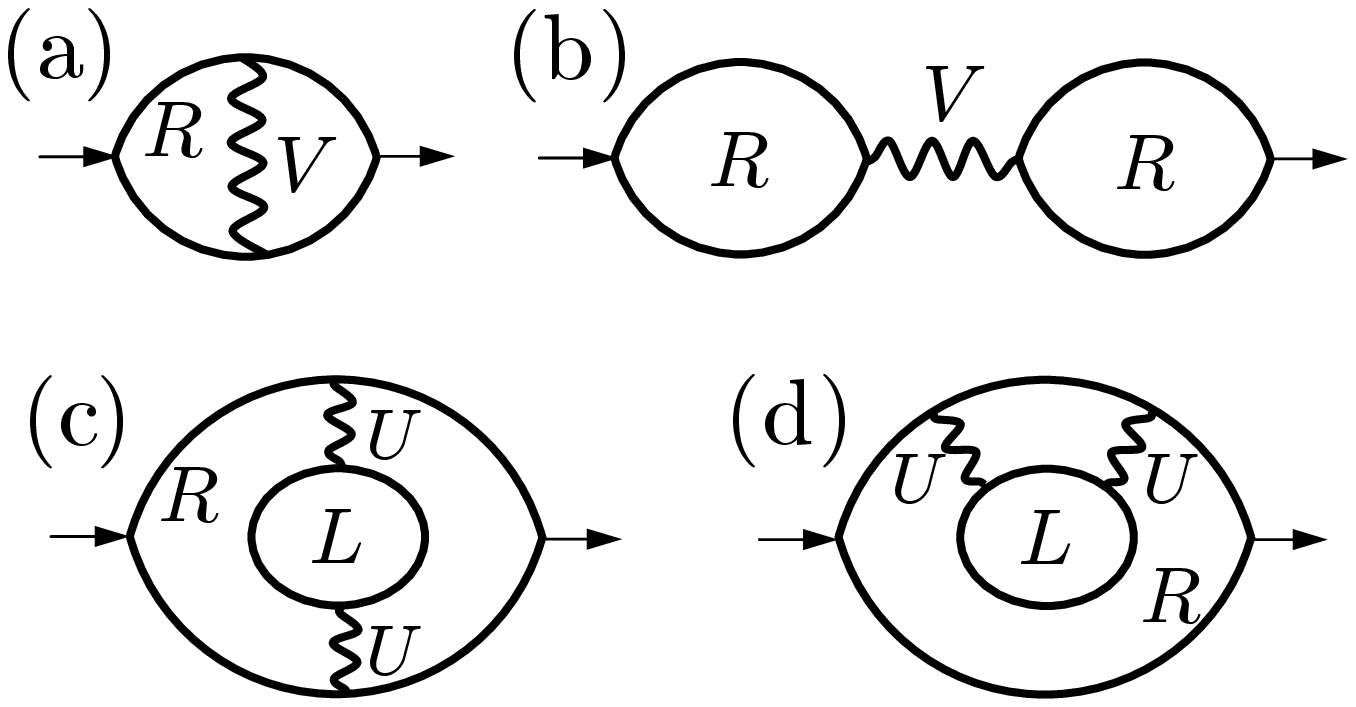}
\caption{ (a),(b) First-order correction to $S(q,\omega)$, logarithmically 
divergent at $\omega\to\omega_-\!+0$ and $\omega\to\omega_+\!-0$. 
Each arc denotes a Green function for right-movers. 
(c),(d) Second-order contributions that diverge logarithmically 
at $\omega\to\omega_+\!+0$; internal loops correspond 
to left-movers.
\label{Fig2} }
\end{figure}

Let us concentrate on the immediate vicinity of the lower absorption
edge,  $\omega\to\omega_-$. Divergent corrections to $S(q,\omega)$
appear already in the first order in the interaction strength, see the
diagrams (a) and (b) in Fig. \ref{Fig2}. Evaluation of these contributions
(which differ from zero only at $\omega_-<\omega<\omega_+$) yields
\beq
\frac{\delta S(q,\omega)}{S_0(q)}
= \mu\ln\!\left[\frac{\delta\omega~}{\omega-\omega_-\!}\right],
\quad
0<\omega-\omega_-\ll\delta\omega
\label{11}
\eeq
with $\mu$ given by \Eq{6}. 
It is not difficult to pinpoint the origin of the logarithmic divergencies
in the perturbation theory. Absorption of a photon results in the creation
of a ``deep'' hole at $k\approx -q$. Particles near the Fermi level may
then scatter off the hole with a small $(\ll q)$ momentum transfer.
Excitation of multiple low-energy particle-hole pairs then leads to the
power-law enhancement of the absorption rate, similar to the edge
singularity in the X-ray absorption spectra in metals~\cite{Mahan}.
Unlike  the conventional X-ray singularity problem, in our case the
deep hole is mobile.  It is known, however, that in 1D the edge singularity
remains intact even when the dynamics of the hole is taken into
account~\cite{Furusaki}.

The  above perturbation theory analysis and the analogy with the
X-ray singularity suggest that the most divergent terms of the
perturbative expansion can be summed by replacing the original
model \Eq{10} with an appropriate effective  Hamiltonian~\cite{Mahan}. 
It should include \textit{two} narrow ($k_0\ll q$) strips of states: one 
around the (right) Fermi point $k=0$, and another around $k=-q$. 
The former accommodates low-energy particle-hole pairs, while the 
later hosts a ``deep'' hole. The corresponding effective Hamiltonian 
$H_-$ is then obtained by projecting \Eq{10} onto the states of 
right-movers with $|k|<k_0\ll q$ and $|k+q|<k_0$ (the $r$ and 
$d$-subbands in Fig.~\ref{3}), while states outside these intervals 
are regarded as either empty or occupied.

\begin{figure}[h]
\includegraphics[width=0.3\columnwidth]{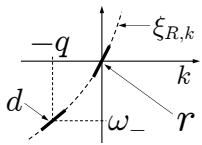}
\caption{
The states of right-movers included in the effective Hamiltonian \eq{12}.
\label{Fig3}
}
\end{figure}

Furthermore, for $k_0\ll q$ the spectrum within the two subbands can 
be linearized. Using
\[
\psi_r (x) = \!\!\sum_{|k|<k_0}\!\frac{e^{ikx}}{\sqrt{L}}\,\psi_{R,k},
\quad
\psi_d(x) = \!\!\!\sum_{|k+q|<k_0}\!\!\!\!\frac{e^{i(k+q)x}}{\sqrt{L}}\,\psi_{R,k},
\]
we write the projected Hamiltonian in the coordinate representation~\cite{footnote3},
\beqa
H_- \!&\!=\!&\!
\int\!dx\,\psi^\dagger_r\bigl(-iu_0\partial_x\bigr)\psi^\pdag_r
+ \!\int\!dx\, \psi_d^\dagger\bigl(-\omega_-\! - iu_{-q\,}\partial_x\bigr)\psi_d^\pdag
\nn\\
&\,&\qquad-~(V_0-V_q)\!\!\int\!dx\,\rho_d(x)\rho_r(x). 
\label{12}
\eeqa 
Here $\rho_r(x) = \,\colon\!\psi^\dagger_r(x)\psi^\pdag_r(x)\colon$ 
and $\rho_d(x) = \psi^\pdag_d(x)\psi^\dagger_d(x)$ are densities 
of particles and holes in the corresponding subbands (the colons 
denote the normal ordering). The velocities $u_0$ and $u_{-q}$ 
are given by 
\beq 
u_p = v + V_0/2\pi + p/m 
\label{13} 
\eeq 
with $p=0,-q$, and include corrections due to both the interaction 
and the spectrum nonlinearity (we neglected $V_p-V_0\propto p^2$ 
here). Finally, $\omega_-$ is the lower absorption edge given by \Eq{3} 
with $u=u_0$.

In terms of the effective Hamiltonian \eq{12}, the structure factor
\Eq{1} takes the form
\beq
S(q,\omega)\! = \!\int\!dx\,dt\,e^{i\omega t}
\bigl\langle B(x,t)B^\dagger(0,0)\bigr\rangle.
\label{14}
\eeq
The operator $B^\dagger= \psi^\dagger_r\psi^\pdag_d$ creates an
inter-subband particle-hole pair (an exciton). With the effective 
Hamiltonian \eq{12}, the correlation function \eq{14} can be 
evaluated by known methods~\cite{Schotte,Furusaki}. Indeed, the 
total number of $d$-holes $N_d = \int\!dx\,\rho_d(x)$ commutes 
with $H_-$. Since the entire $d$-subband lies below the Fermi level, 
the ground state of $H_-$ corresponds to $N_d=0$. The operator 
$B^\dagger$ in \Eq{14} creates one $d$-hole, which propagates 
until it is annihilated by the operator $B$. Therefore, as far as the 
evaluation of \Eq{14} is concerned, $H_-$ can be simplified even 
further by replacing $\psi_d(x)\to{\mathcal P}\psi_d(x){\mathcal P}$, 
where ${\mathcal P}$ is a projector onto states with $N_d=0,1$.
It is easy to see that the projected operators satisfy
\[
\psi_d(y)\rho_d(x) = 0,
\quad
\rho_d(x)\psi_d(y) = \delta(x-y)\psi_d(y),
\]
from which it follows that $\bigl[\rho_d(x),\rho_d(y)\bigr] =0$.

We now bosonize $\psi_r$-field in \eq{12} according to~\cite{Giamarchi}
\beq
\psi_r(x) = \sqrt{k_0}\,e^{i\varphi(x)},
~~
\bigl[\varphi(x),\varphi(y)\bigr]
= i\pi\sign(x-y),
\label{15}
\eeq
which yields $H_- = H_0 + \delta H$, where
\beqa
H_0 &\!=\!&\!\! \frac{u_0}{4\pi}\!\int\!dx\,(\partial_x\varphi)^2
+ \int\!dx\, \psi_d^\dagger\bigl(-\omega_-\! - iu_{-q\,}\partial_x\bigr)\psi_d^\pdag,
\nn \\
\delta H &\!=\!&- \frac{1}{2\pi}\,(V_0-V_q)\!\!\int\!dx\,\rho_d(x)\,\partial_x\varphi.
\label{16}
\eeqa
The Hamiltonian $H_-$ can be diagonalized by the unitary 
transformation~\cite{Schotte} $\widetilde H_- = e^{iW}H_- e^{-iW}$ 
with
\[
W = \theta\!\!\int\!dx\,\rho_d(x)\,\partial_x\varphi, 
\quad 
\theta = \frac{1}{2\pi}\,\frac{V_0-V_q}{u_{-q}-u_0} = - \frac{\mu}{2},
\]
where $\mu$  is given by \Eq{6}. To the linear order in $V$, the 
transformation yields $\widetilde H_- = H_0$. At the same time, 
the transformation modifies the operator $B^\dagger$,
\beq
\widetilde B^\dagger(x)
= e^{iW}B^\dagger(x)\, e^{-iW}
= \sqrt{k_0}\,e^{-i(\!1-\mu/2)\varphi}\psi_d\,.
\label{18}
\eeq
Since $\widetilde H_- = H_0$ is quadratic, evaluation of \Eq{14} is 
straightforward. The structure factor vanishes identically at
$\omega<\omega_-$, while at $\omega>\omega_-$ it is given by \Eq{5},
valid with logarithmic accuracy [i.e., up to a numerical factor in the square 
brackets in \Eq{5}].

Similar procedure can be employed to calculate the structure factor near
the upper edge $\omega=\omega_+$. At $\omega\to\omega_+-0$ 
the first-order in $V$ correction to $S(q,\omega)$ [see Fig.~\ref{Fig1}(a,b)]
diverges,
\beq
\frac{\delta S(q,\omega)}{S_0(q)}
= \mu\ln\!\left[\frac{\omega_+\!-\omega}{\delta\omega}\right],
\quad
0<\omega_+\!-\omega\ll\delta\omega.
\label{19}
\eeq

The non-vanishing contributions to $S(q,\omega)$ at
$\omega>\omega_+$ appear in the second order in $U$, see
diagrams (c) and (d) in Fig.~\ref{2}.  These diagrams describe a
process in which an absorption of a photon results in the final
state that has two particle-hole pairs on the opposite branches of
the Fermi surface \cite{drag}. The corresponding contribution
reads 
\beq 
\frac{\delta S(q,\omega)}{S_0(q)} 
= \nu\ln\!\left(\!1 + \frac{\delta\omega~}{\omega-\omega_+\!}\right)
\!\frac{2_{}\omega_+}{\omega+\omega_+\!}\,, \quad \omega>\omega_+
\label{20} 
\eeq 
with $\nu$ given by \Eq{8}.
\Eq{20} reduces to \Eq{9} at $\omega-\omega_+\gg\delta\omega$, and
diverges logarithmically at $\omega\to\omega_+$.

As above, the divergent contributions can be summed up by
replacing the original model \Eq{10} with an appropriate effective
Hamiltonian. In this case the $d$-subband lies well above the
Fermi level (near $k=q$), and contains at most a single
high-energy particle. However, unlike at $\omega\to\omega_-$, the
interaction with the left-movers now has to be explicitly taken
into account~\cite{lower_edge}. The counterpart of \Eq{12} then reads 
\beqa 
H_+ \!\!&\!=\!&
\!\!\int\!dx\Bigl\lbrace iu_0
\bigl(\psi^\dagger_l\partial_x^\pdag\psi^\pdag_l\!
-\psi^\dagger_r\partial_x^\pdag\psi^\pdag_r\bigr) +
\psi_d^\dagger\bigl(\omega_+\! -iu_{q\,}\partial_x^\pdag\!\bigr)\psi_d^\pdag 
\Bigr\rbrace
\nn \\
&\,&+ \int\!dx\Bigl\lbrace
(V_0-V_q)\rho_d\rho_r
+ U_0(\rho_r + \rho_d)\rho_l
\Bigr\rbrace,
\label{21}
\eeqa
where the field operators are defined similarly to $\psi_{r,d}$ in 
\Eq{12}, and $\rho_d,\rho_r$, and $\rho_l$ are the normal-ordered 
particle densities. The structure factor is given by \Eq{14} with the 
exciton creation operator $B^\dagger = \psi^\dagger_{d\,}\psi^\pdag_r$. 
Similar to the above, after bosonizing the $r$ and $l$ subbands, the 
effective Hamiltonian $H_+$ can be diagonalized by means of a unitary 
transformation. A straightforward, although lengthy calculation~\cite{unpub} 
then yields \Eq{7} for the structure factor. Note that at $\omega\to\omega_+$ 
the edge singularity leads to the \textit{suppression} of the  absorption rate. 
Technically, the difference with \Eq{5} arises because, unlike $d$-holes 
in \Eq{12}, $d$-particles in \Eq{21} move \textit{faster} than the 
particles near the Fermi level: $u_q-u_0=q/m>0$.

The above consideration, leading to power-law singularities in
$S(q,\omega)$, is limited to zero temperature $T=0$. However, as long
as temperature remains low, $T\ll\delta\omega$, its main effect is to
cut off the power-law singularities. This amounts to the replacement
$|\omega-\omega_\pm|\to \max\bigl\lbrace\pi
T,|\omega-\omega_\pm|\bigr\rbrace$ in Eqs. \eq{5} and \eq{7}.

In conclusion, we found the dynamic  structure factor $S(q,\omega)$ 
of interacting fermions in one dimension, without resorting to the TL 
model. This allowed us to uncover certain universal features in the 
behavior of $S(q,\omega)$. The structure factor has a threshold, 
$\omega=\omega_-(q)$, stemming from kinematic constraints,
see Fig~\ref{Fig1}. The divergence of $S$ at the threshold, 
\Eq{5}, is characterized by exponent $\mu$, which is a smooth 
function of the wave number $q$. For weak interactions, the explicit 
form of the function $\mu(q)$, valid at any $q<2p_F$, is given 
in \Eq{6}. We believe however, that the appearance of the power-law 
divergency in $S$ along the entire boundary $\omega=\omega_-(q)$ 
is a generic feature, not limited to weak interactions. Indeed, the methods
we employed to arrive at \Eq{5} rely on smallness of $\mu(q)$, which 
is achieved at sufficiently small $q$ at any strength of interactions $V$ 
and $U$. Also, at $q\to 2p_F$ the power-law divergence of $S$ is 
evident from the conventional Luttinger liquid 
theory~\cite{Giamarchi}. Remarkably, that the origin of the threshold 
singularity in $S$ can be traced back to the physics of Mahan 
exciton in theory of X-ray edge singularity~\cite{Mahan}, and remains 
universal at any $q$. The same physics dictates the existence of a 
power-law singularity in $S(q,\omega)$ at $\omega=\omega_{+}(q)$, 
see 
\Eq{7}. 

Besides being of a fundamental interest, the knowledge of the
structure factor with nonlinear dispersion relation is crucial for
understanding the Coulomb drag, photovoltaic effect, and other
phenomena that owe their existence to the particle-hole asymmetry.
The developed theory is also applicable to the inelastic neutron
scattering off antiferromagnetic spin chains~\cite{Giamarchi,muller}. 

\begin{acknowledgments}
We are grateful to I.~Affleck, F.~Essler, and P.~Wiegmann for useful
discussions. This project was supported by NSF grants DMR02-37296,
DMR-0405212, EIA02-10736, and by A.~P.~Sloan foundation.
\end{acknowledgments}


\begin{thebibliography}{0}
\bibitem{fdt} The structure factor \Eq{1} is related to the
imaginary part of the corresponding retarded correlation function 
by the Fluctuation-Dissipation Theorem, see~\cite{Nozieres}.
\bibitem{width}
In the bosonic representation, even the simple result
$\delta\omega\sim q^2\!/m$ is non-perturbative: the first-order
in $1/m$ correction to $S$ vanishes, while the next
one diverges~\cite{Samokhin}.
\bibitem{footnote3}  The projection transforms the interaction term in the
Hamiltonian Eq.~(\ref{10}) into the form
$V_0\psi^\dagger_d\psi^\pdag_d\psi^\dagger_r\psi^\pdag_r+
V_q\psi^\dagger_d\psi^\pdag_r\psi^\dagger_r\psi^\pdag_d$, which
after the normal ordering yields the last term in Eq.~(\ref{12}).
\bibitem{lower_edge}
Inclusion of the interaction with the left-movers in the effective
Hamiltonian \Eq{12} would lead to a second order in $U$ correction
to the exponent $\mu$~\cite{unpub}.
\end{thebibliography}
\end{document}